\documentclass[twocolumn,prl,showpacs]{revtex4}
\usepackage{graphicx}%
\usepackage{dcolumn}
\usepackage{amsmath}
\usepackage{latexsym}

\begin{document}
\title{Free Energy of Twisted Semiflexible Polymers}
\author{Supurna Sinha}
\affiliation{Raman Research Institute, Bangalore 560 080, India.\\}
\begin{abstract}
We investigate the role of fluctuations in 
single molecule measurements of torque-link ($t-lk$)
curves. For semiflexible polymers of finite persistence length
( i.e. polymers with contour length $L$ comparable to the
persistence length $L_P$), the 
torque versus link 
curve 
in the constant torque (isotorque)
ensemble is distinct from the one in the 
constant link (isolink) ensemble. 
Thus, one encounters the conceptually interesting issue of a 
``free energy of transition'' in switching ensembles while making
torque-link measurements. We predict the dependence on the 
semiflexibility parameter $\beta = L/L_P$ of this extra
contribution to the free energy which shows up as an area in the 
torque-link plane. This can be tested against future torque-link
experiments with single biopolymers.
We bring out the inequivalence of torque-link curves for a stiff 
polymer and present explicit analytical expressions for the 
{\it distinct} torque-link relations in the two ensembles 
and the free energy difference in switching ensembles in this 
context. The predictions of our work
can be tested against
single molecule experiments on torsionally constrained biopolymers.
\end{abstract}
\maketitle


Statistical mechanics of semiflexible polymers is of great current 
interest. Research in this area has been motivated by
experiments\cite{bust,yuri} on biopolymers in which single molecules are
stretched and twisted to measure elastic properties. These
experiments are designed to understand the role of semiflexible polymer
elasticity\cite{twist,writhe,dnael} in, for instance, the 
packaging of 
these polymers in a cell nucleus. 
The process of DNA
transcription can generate supercoiling. It is also regulated by 
supercoiling\cite{Strick}.
In a typical experiment\cite{Strick} probing the twist elasticity of a DNA 
molecule,
the ends of a single molecule of double stranded DNA are attached to
a glass plate and a magnetic bead.
Magnetic fields are used to rotate
the bead and magnetic field gradients to apply forces on the bead.
By such techniques the molecule is stretched and twisted and the extension
of the molecule is monitored by the location of the
bead. One can thus measure the extension
of the molecule as a result of the applied link and force \cite{Strick}
and also make a measurement of the torque versus 
applied link \cite{bryant} .

In the context of force-extension measurements, an 
isometric setup is described by
the  
Helmholtz free energy, whereas an isotensional setup is
described by the Gibbs free energy\cite{landau}. 
In statistical mechanics, 
these two ensembles are distinct\cite{sinha,mar,keller}.
In the thermodynamic
limit
these
two descriptions agree, but semiflexible polymers (those with contour
length $L$ comparable to the persistence length $L_P$, i.e. 
$\beta=L/L_P \simeq 1$) are {\it not} at the
thermodynamic limit.

In the present paper we explore
this issue in the context of torque-link measurements. The two 
distinct statistical mechanical ensembles here are 
the constant torque ensemble (isotorque ensemble) and the constant link 
ensemble (isolink ensemble).                       
Here we focus on the role of 
fluctuations in single molecule torque-link experiments.
In order to correctly interpret such experiments one needs to 
understand the effect of fluctuations on the measured quantities. 
For instance, it turns out, that an experiment in which the 
link applied to a polymer molecule is fixed (isolink) and the 
torque fluctuates yields a different
result from one in which the torque is held fixed 
(isotorque) and 
the link fluctuates\cite{bouchiat}. 
This difference can be traced to 
large fluctuations about the mean value of the torque or the link,
depending on the experimental setup.
Experimentally, both isolink and isotorque
ensembles are realizable.
Here is a schematic description of an experimental setup 
for torque-link measurements of single biopolymer molecules. A polymer
molecule attached to a glass plate on one end is suspended in a 
suitable medium with a 
magnetized bead attached to the other end. The magnetized bead is kept
in a magnetic trap. 
One can realize an isolink setup by using a 
``stiff magnetic trap'' and an isotorque setup by 
using a ``soft magnetic trap'' which 
allows the applied link to fluctuate but applies a fixed torque
to the molecule\cite{laporta,torquefoot}. 
The fluctuations in torque-link measurements
vanish only in the thermodynamic limit of very long polymers. 
In the next section we describe the setup in more detail.
                                                                           
The paper is organized as follows. In Sec. $II$ we discuss the isolink
and the isotorque ensembles. 
In Sec. $III$ we illustrate the phenomenon of inequivalence of ensembles 
in torque-link measurements by explicitly presenting 
analytical expressions in the context of a stiff polymer. 
In Sec. $IV$ we draw attention to the notion of  
the free energy of transition in going from one ensemble to 
another and its dependence on 
semiflexibility which can be 
tested against future single molecule 
experiments on torsionally constrained polymers
and simulations. 
Finally, we conclude the paper in Sec. $V$.


Consider an experiment, as described in the Introduction, 
in which one end of a biopolymer molecule 
is attached to a glass plate and the other end is attached to a 
magnetized bead (of magnetic moment ${\vec \mu}$)
kept in a magnetic field ${\vec B}$ which is used to 
rotate the bead. 
We suppose both ${\vec B}$ and ${\vec \mu}$ are parallel to the
glass plate.
The energy of the bead is given by 
\begin{equation}
E = -{\vec \mu}.{\vec B}={-\mu B \cos(\theta-\theta_0)}
\label{trap}
\end{equation}
$\theta_0$ describes  the direction of the magnetic field and 
$\theta$ the direction of the magnetic moment ${\vec \mu}$.
The variables $lk$ and $lk_0$, which keep track of the number $n$ of turns
of the bead are 
related to $\theta$ and $\theta_0$
as follows:
$lk=\theta+2n\pi$ and $lk_0=\theta_0+2n\pi$.
Consider $P(lk)d{lk}$, the number of 
configurations (counted with 
Boltzmann weight) for a polymer of contour length $L$ and bend 
persistence length $L_P$, characterized by a semiflexibility
parameter $\beta=\frac{L}{L_P}$, 
in a link  
interval $d lk$ of $lk$. 
The free energy defined by ${\Phi}(lk)= -\frac{1}{\beta}ln P(lk)$ 
is the free energy pertaining to a
fixed link.
The partition function\cite{keller,partfoot} for the combined system 
consisting 
of the polymer molecule
{\it and} the magnetized bead in the magnetic field is given by:
\begin{equation}
Z(lk_0,\beta)=\int_{-\infty}^{+\infty}{d lk e^{-\beta 
{\Phi}(lk)}e^{{\beta \mu B}\cos(lk
-lk_0)}}
\label{part}
\end{equation}
 
{\underline{Constant Link Ensemble: The Limit of a Stiff Trap}}

In the limit of a stiff trap ($\mu B \rightarrow \infty$),
${\vec \mu}$ follows ${\vec B}$ closely  
and 
$lk\approx lk_0$. Thus $\cos(lk - lk_0)\approx 1-\frac{(lk - lk_0)^2}{2}$.
In this limit the partition function for the combined system 
(molecule+trap) reduces to: 
\begin{equation}
Z(lk_0,\beta)={e^{\beta \mu B}}\int_{-\infty}^{+\infty}{d lk e^{-\beta 
{\phi}(lk)}e^{-\beta \mu B\frac{(lk
-lk_0)^2}{2}}}
\label{party}
\end{equation}
Clearly, in this limit the Gaussian factor pertaining to the magnetic trap
approaches a delta function and we have:
\begin{equation}
Z(lk,\beta) \approx e^{-\beta {\Phi}(lk)}
\label{stifftrap}
\end{equation} 
Here we have switched notation to write $lk$ in place of $lk_0$.
Thus a stiff trap realizes the
Constant Link (isolink) ensemble by constraining fluctuations in $lk$ .
In order to change the applied link from $lk$ to $lk+dlk$ one  
applies a torque $<t> = \frac{\partial{\Phi}}{\partial{lk}}$.
Thus one gets a torque-link $(<t>,lk)$ curve by plotting 
$<t>$ versus $lk$. 

{\underline{Constant Torque Ensemble: The Limit of a Soft Trap}} 

In the opposite limit of a soft trap $\mu B$ is small but 
large enough that the polymer does not get untwisted.   
In such a situation the link fluctuates. One can adjust 
$lk_0$ such that $lk-lk_0 \approx \pi/2$. The magnitude of the torque
$|{\vec t}|=|{\vec \mu}\times{\vec B}| = t = \mu B$ is held fixed for a 
particular measurement.
In this limit the torque $t$ is the control parameter which can be 
changed from one reading to the next by changing the magnitude $B$
of the magnetic field. 
A feedback loop is used to ensure that $<lk - lk_0>$ is maintained at
$\pi/2$. Clearly, the potential energy for the 
trap, on expanding 
around $lk-lk_0=\pi/2$ and retaining terms to linear
order in $lk-lk_0$ takes the form: 
\begin{equation}
E = -{\vec \mu}.{\vec B}={\mu B (lk-lk_0)}
=t (lk - lk_0)
\label{trap}
\end{equation}

Thus, in this limit Eq. ($\ref{part}$) gives   
the following expression for the partition function 
${\tilde Z}(t,\beta)=Z(lk_0,\beta)e^{\beta t lk_0}$
(where $lk_0$ is determined by the condition 
$<lk - lk_0> = \frac{\pi}{2}$)
for the combined system
consisting of the polymer molecule and the trap \cite{keller,partfoot}:
\begin{equation}
{\tilde Z}(t,\beta)=\int_{-\infty}^{+\infty}{dlk e^{-\beta 
\Phi(lk)}e^{\beta 
tlk}}
\label{softrap}
\end{equation}
Thus in a soft trap link $lk$ fluctuates but 
torque fluctuations are constrained\cite{kreuz}. One thus realizes the 
Constant Torque (isotorque) ensemble.
${\tilde Z}(t)$\cite{footz} is the generating function for the $lk$ 
distribution.
Given the constant torque free energy 
$\Gamma(t)={-\frac{1}{\beta}}ln{\tilde{Z}}(t)$ 
one gets the mean link $<lk>=-\frac{\partial \Gamma}{\partial t}$
and the $(t,<lk>)$ torque-link relation.

Notice that ${\tilde Z}(t)$ is the Laplace transform 
of 
$Z(lk)$.    
In the thermodynamic limit of long polymers ($\beta 
\rightarrow \infty$) the Laplace transform integral Eq.(\ref{softrap}) 
is dominated by 
the saddle point value and therefore  
${\Phi}(lk)$ and $\Gamma(t)$ are related by a Legendre transform:
\begin{equation}
{\Phi}(lk) = \Gamma(t) +tlk.
\label{legendre}
\end{equation}
For finite $\beta$ i.e. for a polymer of finite extent, 
the saddle point approximation 
no longer holds true and fluctuations about the saddle point value
of the free energy become important.  
Thus one finds that for a finite $\beta$, ${\Phi}(lk)$ and $\Gamma(t)$ are 
{\it not} 
Legendre transforms of each other
but are related via a 
Laplace transform [Eq. (\ref{softrap})]. 



We illustrate the issue of inequivalence of ensembles in the 
context of torque-link measurements explicitly by considering an 
analytically 
tractable and instructive special case, the torque-link relation for 
a stiff polymer. 

Our starting point is the WLC Hamiltonian with bend and twist degrees of
freedom in the presence of a stretching force $f$ and a torque $t$
\cite{bouchiat}:
\begin{equation}
H = p_\theta ^2/2 + (p_\phi - A_\phi)^2/2 \sin^2 \theta -f\cos\theta  
-\alpha t^2/2                        
\label{finalhamilton}
\end{equation} 
Here $p_\theta$ and $p_\phi$ are momenta conjugate to the Euler angles
$\theta$ and $\phi$. The momentum conjugate to the Euler angle
$\psi$ is $p_\psi = it$, a constant of motion, which contributes a term
$-\alpha t^2/2$ to the Hamiltonian where 

$\alpha$ is the ratio of the
bend persistence length $L_{P}$ to the twist persistence length
$L_{T}$. 
The `vector potential'
$A_{\phi}={it(1-\cos{\theta})}$\cite{writhe,bouchiat}.
For a stiff polymer with one end clamped along the 
${\hat z}$ direction, we can approximate the sphere
of directions by a tangent plane
at the north pole of the sphere as the
angular coordinate $\theta$ always remains small.
In this limit (the PWLC model \cite{writhe})
where the tangent vector never 
wanders too far away from the 
north
 pole of the sphere of directions,
the polymer Hamiltonian\cite{twist,writhe,bouchiat,nelson} reduces 
to :
$$H_{PWLC}=\frac{p_{\theta}^2}{2}+\frac{(p_{\phi}-A_{\phi})^2}{2{\theta}^2}
-\frac{\alpha t^2}{2}-f(1-\frac{{\theta}^2}{2})$$
$$=H_P -f -\frac{\alpha t^2}{2}$$
where $H_P$ is the Hamiltonian of interest in the paraxial limit 
after we take out a constant piece. 
In this limit $A_{\phi}=\frac{it\theta^2}{2}$.
We introduce Cartesian coordinates $\xi_{1}=\theta \cos{\phi}$ and
$\xi_{2}=\theta \sin{\phi}$
on the tangent plane $R^2$ at the north pole.

The PWLC model has been applied earlier in the context of flexible 
polymers at high tension\cite{writhe}. Here we apply it in the 
context of stiff polymers at $f=0$.
In the stiff limit, the tangent vector to the polymer points essentially
along a fixed direction (the north pole) even at $f=0$. 
As in Ref. \cite{writhe} we restrict to the case of 
$\alpha = 0$\cite{footwr}.

In Ref. \cite{stiff} we have derived the partition function 
$Z(f)$
of a stiff polymer with both end tangent vectors pointing along
a fixed direction ${\hat z}$ at a
force $f$.
In the presence of a force $f$ {\it and} a torque $t$ this expression 
goes over to: 
\begin{equation}
{\tilde Z}^f(t) = \sqrt{f-t^2/4}\exp (\beta  f)/(2 \pi \sinh\big[\beta 
\sqrt{f-t^2/4}] \big).
\label{zofft}
\end{equation}
Notice that the ``effective force'' $\sqrt{(f - t^2/4)}$ 
replaces $f$ when there is a competition between a force $f$ and a torque 
$t$. This is simply due to the fact that the Hamiltonian $H_P$ in the 
presence of a force $f$ and a torque $t$ pertains to that of a 
two dimensional harmonic oscillator with a frequency 
$\omega =\sqrt{(f - t^2/4)}$ rather than $\omega =\sqrt{f}$ which is 
the expression for the frequency of a two dimensional harmonic
oscillator in the presence of a force $f$ and {\it no} torque. 

At zero force, the expression simplifies and reduces to:
\begin{equation}
{\tilde Z}(t) = t/(4 \pi \sin\big[\beta 
t/2] \big).
\label{zofot}
\end{equation}
>From the partition function ${\tilde Z}(t)$, we get the following 
analytical 
expression for the torque-link $(t,<lk>)$ relation in the isotorque 
ensemble:
\begin{equation}
<lk> = \big(\frac{1}{\beta t} -\frac{1}{2}\cot\big[\frac{\beta t}{2}\big] 
\big).
\label{lkt}
\end{equation}
 
Given ${\tilde Z}(t)$ one can get the partition function $Z(lk)$
in the conjugate domain of an isolink ensemble (See Sec. $II$)
\cite{stiff,BMM}:
\begin{equation}
 Z(lk) = (\pi/2\beta)/\big(\cosh\big[ 
\pi lk] \big)^2.
\label{zofflk}
\end{equation}
This leads to the following torque-link $(<t>,lk)$ relation:
\begin{equation}
<t> = (2\pi/\beta)\tanh\big[ 
\pi lk].
\label{tlk}
\end{equation}

Thus, the torque-link $(t,<lk>)$ relation obtained in the isotorque
ensemble (Eq.[\ref{lkt}]) is distinct from the one $(<t>,lk)$ 
obtained in the isolink ensemble (Eq.[\ref{tlk}]) 
[See Fig. $1$]. 

\begin{figure}[t]
\begin{center}
\includegraphics[width=6.0cm]{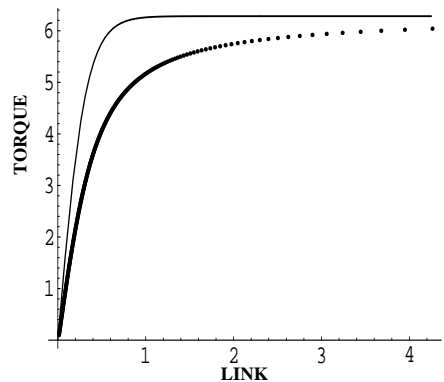}
\end{center}
\caption{Torque-Link Curve in the isolink (upper curve) and isotorque 
(lower curve) ensembles for $\beta=1.$ }
\label{torcontrast}
\end{figure}

In the next section we derive explicit expressions for the 
extra contribution to the free energy due to changes in ensemble. 


In an isotorque setup, torque is the control parameter and one 
measures the mean link to plot the torque-link ($t, <lk>$) 
curve.
In an isolink setup the roles of link and torque are interchanged.
Consider going from a small torque $(lk_{1}, t_1)$ to a large torque 
$(lk_{2}, t_2)$ configuration via an 
isotorque setup and returning from a large link $(lk_{2}, t_2)$ to 
a small link $(lk_{1}, t_1)$ configuration via an isolink setup. 
Since the torque-link relation depends on the
chosen ensemble, in general there will be two distinct curves 
joining the points $(lk_{1}, t_1)$ and $(lk_{2}, t_2)$ in the 
torque-link plane, describing the two processes.
Such a transformation could lead to a net area being enclosed in the
torque-link plane. 
This appears paradoxical since it seems to suggest that one 
can extract work from the system via a cyclic process. 
The resolution of this paradox is as follows. 
In completing the cycle and returning to the initial
state one is in fact changing ensembles twice at the two end points.
Such {\it ensemble changes} in a cyclic transformation involve {\it 
finite changes} in free energy which need to be taken into consideration. 
In particular, we notice for the special case of a stiff polymer the
torque-link relations (Eq.[\ref{lkt}] and Eq. [\ref{tlk}]) lead to a 
{\it free energy difference} of :
$\Delta_{stiff} =$ 
$${\frac{2}{\beta}}ln|\frac{\cosh(\pi lk_2)}{\cosh(\pi lk_1)}|
- (t_2 lk_2 - t_1 lk_1) 
+\frac{1}{\beta}\bigg[ln|\frac{t_2}{t_1}|
-ln|\frac{\sin(\frac{\beta t_2}{2})}{\sin(\frac{\beta t_1}{2})}|\bigg] $$
for a transformation between the states
$(lk_{1}, t_1)$ and $(lk_{2}, t_2)$.  
In the stiff regime (i.e. at small $\beta$) the dependence of the 
free energy on $\beta$ will be dominated by the first term.
In other words, the free energy difference 
$\Delta_{stiff} \sim \frac{1}{\beta}$. 
This is a prediction of our
analysis which can be tested against experiments with stiff biopolymers
like actin filaments. 

An analysis similar to the one in Ref. \cite{sinha}
applied to the context of torque-link measurement shows that one gets
%
a contribution 
$\Delta = \frac{1}{2\beta}ln{{\Phi}^{''}(lk_{*})}$
to the free energy coming from fluctuations 
around the 
long polymer ($\beta \rightarrow \infty$) limit by expanding
$\Delta({lk}) = {\Phi}(lk)-t lk$ around the saddle point value
$lk=lk_{*}$ pertaining to the long polymer limit. Here 
${\Phi}(lk)$ corresponds to the isolink free energy.  
This extra contribution $\Delta({lk})$ vanishes in the 
limit of $\beta \rightarrow \infty$.
For finite $\beta$, this nonzero contribution 
to the free energy accounts for the transition between the constant
link ensemble and the constant torque ensemble.
Work is done on the bead by the trap in making the trap stiffer, while in 
going from a stiff trap to a soft 
trap work is extracted from the bead by the trap. The net work done is the 
difference between the work done at the two ends
of the torque-link curves in switching ensembles.
This net work exactly cancels out the nonzero 
area enclosed in the torque-link plane.  
The net area enclosed in the torque-link plane pertaining
to the ``Free Energy of Transition'' scales as $1/\beta$ and therefore
grows with the rigidity of the polymer. 
These predictions of our
study can be tested against future simulations and 
single molecule experiments.


To summarize, in this paper we have studied inequivalence of ensembles
for torque-link measurements. We have calculated the free energy 
difference between torque-link measurements in the isotorque and 
isolink ensembles for a stiff polymer. In addition, we have 
determined the contribution to the 
``free energy of transition'' in going between the isolink 
and the isotorque ensembles
by expanding the free energy difference $\Delta({lk})$ around the long 
polymer limit.
We predict the dependence on the 
semiflexibility parameter $\beta$ of this extra
contribution to the free energy which shows up as an area in the 
torque-link plane. 
For the special case of a stiff polymer we find explicit
analytical expressions for the 
torque-link $(t,<lk>)$ relation obtained in the 
isotorque
ensemble (Eq.[\ref{lkt}]) and show that it is distinct from the one 
$(<t>,lk)$ 
obtained in the isolink ensemble (Eq.[\ref{tlk}]). 
We also show that in this stiff regime the free energy difference 
$\Delta_{stiff} \sim \frac{1}{\beta}$.
All the predictions mentioned here can be qualitatively and 
quantitatively tested
against future single molecule experiments on torsionally 
constrained biopolymers.

The theoretical predictions presented in this study are 
expected to 
generate interest in torsionally constrained single molecule experiments 
which 
will eventually lead to a deeper 
understanding of the role of twist 
elasticity in biological processes 
involving gene regulation\cite{CELL}.

{\it Acknowledgements:}
It is a pleasure to thank A. Ghosh and D. Marenduzzo for discussions.


\begin{references}
\bibitem{partfoot} We drop additive constants in the free energies or
equivalently, multiplicative constants in the partition functions.
\bibitem{footz}
We suppress the $\beta$ dependence 
and write ${\tilde Z}(t)$ instead of ${\tilde Z}(t,\beta)$
and ${Z}(lk)$ instead of ${Z}(lk,\beta)$
in our subsequent discussion.
\bibitem{footwr}
There is no loss of generality in choosing $\alpha=0$. One can 
extend the analysis to finite $\alpha$ by multiplication by a 
Gaussian factor in the generating function space as outlined in 
\cite{writhe}.
\end{references}
\end{document}